 \documentclass[final,5p,times,twocolumn,authoryear]{elsarticle}

\usepackage{amssymb}
\usepackage{lipsum}
\usepackage{graphicx}
\usepackage{dcolumn}
\usepackage{bm}
\usepackage{bbm}
\usepackage{amsmath}
\usepackage{epsfig}
\usepackage{indentfirst}
\usepackage{psfrag}
\usepackage{subfigure}
\usepackage{amssymb}
\usepackage{color}
\usepackage{hyperref}
\hypersetup{colorlinks,allcolors=black}
\usepackage{ulem}
\usepackage[T1]{fontenc}
\usepackage{comment}
\usepackage{upgreek}
\usepackage{listings}

\def\ii{{\rm i}}  \def\ee{{\rm e}}
  
\DeclareUnicodeCharacter{202F}{\,}

  \def\red{\color{red}}    

\newcommand{\new}[1]{{\color{blue} #1}}

\newcommand*\Laplace{\mathop{}\!\mathbin\bigtriangleup}

\journal{Ultramicroscopy}

\begin{document}

\begin{frontmatter}



\title{Efficient methods for wave propagation in electron microscopy}


\author[first,second]{Zdeněk Nekula}
\affiliation[first]{organization={Institute of Physical Engineering, Brno University of Technology},
            addressline={Technická 2896/2}, 
            city={Brno},
            postcode={616 69}, 
            country={Czech Republic}}
\affiliation[second]{organization={Central European Institute of Technology, Brno University of Technology},
            addressline={Purkyňova 656/123}, 
            city={Brno},
            postcode={612 00}, 
            country={Czech Republic}}

\author[first,second]{Jakub Bělín}

\author[first,second]{Andrea~Kone\v{c}n\'{a}}

\begin{abstract}
Accurate wave-optical simulation in electron microscopy is severely constrained by the extreme sampling requirements imposed by short wavelengths and relatively large convergence angles. Conventional implementations of the angular spectrum method (ASM) rapidly become computationally intractable, often exceeding realistic memory and time limits. We present two numerical approaches -- the scaling angular spectrum method (SASM) and the no-lensing angular spectrum method (NLASM) -- that systematically reduce the sampling requirements while retaining the essential physics of wave propagation. SASM replaces the original optical system with a scaled equivalent in which lens-induced beam convergence or divergence is reduced, lowering memory usage and computational cost by approximately the square of the scaling factor. NLASM suppresses lensing effects altogether, enabling highly efficient propagation away from focal planes. Benchmarking against the Bluestein (chirp-z) transform reveals that the three methods are complementary and together enable wave-optical simulations of complex electron-optical systems previously considered infeasible. These results establish practical pathways toward routine wave-based modeling in electron microscope design.
\end{abstract}




\end{frontmatter}




\section{Introduction}
\label{introduction}

The angular spectrum method (ASM)~\cite{Goodman2005, Voelz2009, Matsushima2003} is a widely used computational framework for modelling wave propagation in optics, acoustics, and related fields. By decomposing an arbitrary wavefront into a superposition of plane waves, the ASM provides an exact solution for free-space propagation and diffraction. In this representation, the transverse field distribution at a given plane is Fourier-transformed to obtain its angular spectrum, each Fourier component acquires a phase shift determined by its spatial frequency and propagation distance, and the propagated field is reconstructed through an inverse Fourier transform.

When implemented numerically, the ASM typically represents the wavefront on a discrete square grid  
$\mathcal{M}_{\mathrm{real\, space}} \in \mathbb{C}^{N \times N}$. Two well-known issues arise from this discretization.  
First, the Fourier transform implicitly assumes periodic boundary conditions, requiring the field to vanish (or be padded) near the edges to avoid unphysical wrap-around artefacts. Second, the grid supports only a finite spatial-frequency bandwidth. High-angle waves, corresponding to high spatial frequencies, must be sufficiently sampled to prevent aliasing and incorrect propagation directions~\cite{Kozacki2015_CompactSBP}. Consequently, the number of pixels must be large enough to satisfy the Nyquist condition~\cite{Nyquist1924} for the most rapidly varying phase term.

To illustrate the sampling requirement, consider a wavefront modified by a thin lens imposing a phase modulation
\begin{align}
    \Delta\varphi(R) = -\frac{k R^{2}}{2f},
\label{Eq:lenssimple}
\end{align}
where $R=\sqrt{x^{2}+y^{2}}$ is the radial coordinate, $k$ the wavenumber, and $f$ the focal length.  
Demanding that the phase difference between neighbouring pixels (with a distance $\Delta R$) remains below $\pi$, i.e. $|\varphi(R)-\varphi(R+\Delta R)| < \pi$, yields the constraint
\begin{align}
    \Delta R < \frac{\pi f}{k R_{\max}},
\end{align}
where $R_\mathrm{max}$ is the radial span of the grid. For strongly focusing systems or short wavelengths, which is often the case in electron optics~\cite{Kirkland2010_AdvancedEM}, this constraint therefore leads to extremely large arrays. If we consider an example of probe focusing in a scanning transmission electron microscope, we typically have a lens with a focal length $f=1\,$mm, a convergence semi-angle of $100\,$mrad at $200\,$kV acceleration voltage~\cite{Muller2006}, that corresponds to a wavenumber 2.5 rad/pm. Direct ASM simulation of such a wavefront would require an array of order $(1.6\times 10^{7})^2$ complex numbers, corresponding to \(\sim 4\) PB of memory—far beyond the capacity of even the largest single-node memory systems (e.g. 160 TB in HPE's ``The Machine''~\cite{TheMachine2015}). Yet, in practice, the desired output is typically a final probe profile containing orders of magnitude fewer pixels.

A variety of improvements to the ASM have been explored. Boundary-reducing strategies were introduced in Ref.~\cite{Matsushima2009}, curved-surface propagation in Ref.~\cite{Hwang2014}, and pixel-scaling techniques for converging and diverging beams in Refs.~\cite{Heintzmann2023,Abedi:24, Matsushima2010_ShiftedASM, Guo2014_HybridGTFGSPF}. While these methods significantly improve numerical efficiency and flexibility, they do not remove the fundamental sampling constraints. Several works~\cite{Kamal2022, BARTHEL20181, Susi2020, CowleyMoodie1957, Ophus2017_STEMReview} circumvent these limitations by foregoing full wave propagation and directly computing the final probe from aperture functions or point-spread functions (e.g. abTEM (AT) code \cite{Susi2020}). While sufficient for many microscopy users primarily interested in the probe–sample interaction, these approaches do not provide the element-by-element wave propagation necessary for optical system design. Due to the prohibitive computational requirements of full-wave methods, designers often resort to particle-tracing models~\cite{LENCOVA2008315,Geer1997, Dahl2000}, abandoning wave-based treatments entirely. Finite-element approaches are likewise impractical in wave optics because the cell size must remain below the picometre-scale wavelength in all three spatial dimensions~\cite{Oskooi2010_MEEP}.

More recently, efficient full-path diffraction calculations based on the Bluestein (BS) chirp-z transform have been demonstrated, enabling flexible input–output sampling and accurate scalar and vector propagation without FFT-imposed grid restrictions~\cite{Leutenegger2006, Hu2020}. This method delivered excellent results in laser optics and provided a robust and efficient tool for light optics designers. As we show in this work, BS also enables overcoming the extreme sampling requirements of electron microscopes in some specific cases, such as propagation from the objective plane to the focal plane. However, there are still examples in electron microscopy that require extreme sampling even with the BS, as we demonstrate in the following.

Here we introduce a scaling angular spectrum method (SASM) that approximates the original optical system with a rescaled system in which angles arising from lensing are reduced by a controllable scaling factor $\delta$. This significantly relaxes the sampling conditions and lowers the required grid size without compromising the essential physics of propagation. For instance, a scaling factor of 1000 reduces the number of grid points 1000$^2$ times. In other words, it can reduce a petabyte ASM calculation to the order of gigabytes, placing previously infeasible simulations within reach of standard workstations or even laptops. A similar improvement is in the number of required computer operations. The ASM is implemented using 2D FFTs, and each 2D FFT requires $2 N^2\mathrm{log}_2(N)$ computer operations~\cite{CooleyTukey1965}, which determines the computational time. Both required computer memory and computational time drop approximately by $\delta^2$. We further modify SASM into a no-lensing angular spectrum method (NLASM). 

In this article, we first present the analytical theory underlying the proposed methods and then consider several examples of both simpler and more complex electron-optics arrangements. We show that the combination of NLASM and SASM can solve complicated optical architectures containing multiple lenses, apertures, diffraction gratings~\cite{Verbeeck2010, McMorran2011_OAM}, or phase plates~\cite{Shiloh_2019,shiloh2018,Verbeeck2018,Verbeeck2023,Verbeeck2023_01}. By benchmarking the new methodology against ASM and BS, we demonstrate its advantages and validate that it enables full-wave optical simulations of high-angle, short-wavelength beams that were previously computationally infeasible.



\section{Analytical theory}
Propagation of an electron wave function in an electron microscope can be typically described by a relativistically corrected Schrödinger equation \cite{hawkeskasper_book}. In this work, we simulate the propagation of electrons in a free space, between electron-optics elements, where the solution to the Schrödinger equation can be found in the form $\psi(\mathbf{r},t)=\mathcal{B}(\mathbf{r})\exp\{-\ii Et/\hbar\}$. $E$ is the energy of the electron, $\hbar$ is the reduced Planck's constant, $\mathbf{r}=(x,y,z)$, and a function $\mathcal{B}(\mathbf{r})$ can be found as a solution to the so-called time-independent Schrödinger equation.
In a free space, this equation is isomorphic with the Helmholtz equation $\Laplace{}\mathcal{B}(\mathbf{r})+k^2\mathcal{B}(\mathbf{r})=0$, where $k$ is the (relativistically corrected) wave vector of the moving electron.

Given the beam function $b_0(\mathbf{R})=b_0(x,y)$ defined in a plane $z=z_0$, the solution to the Helmholtz equation can be found in a form
\begin{equation}
    \mathcal{B}(\mathbf{r})=\mathcal{F}^{-1}\left\{\mathcal{F}\{b_0(\mathbf{R})\}\,\ee^{\ii K_z\Delta z} \right\},
    \label{eqn:FSPM}
\end{equation}
where symbols $\mathcal{F}$ and $\mathcal{F}^{-1}$ refer to the forward and inverse Fourier transform, $\Delta z=z-z_0$ is the propagation distance, and $K_z$ is the longitudinal component of the wavevector.

We note that Eq.~(\ref{eqn:FSPM}) is the basis of the well-known ASM, which relies on its numerical solution. This method is analytically accurate and therefore works even for a non-paraxial case.
However, analytical calculations demonstrating principles on which our proposed  SASM is based, introduced below, can be performed only if the paraxial approximation is applied.
We will therefore demonstrate the fundamentals of SASM on calculations in the paraxial regime.
Then, beam propagation is equivalent to a convolution of the beam function $b_{\mathrm{j}}(\mathbf{R})$, defined in plane $z=z_{\mathrm{j}}$, and the Fresnel propagator $p(\mathbf{R},\Delta z)$, given by a formula 
\begin{equation}
    p(\mathbf{R},\Delta z)=\frac{-\ii k}{2\pi\Delta z}\ee^{\ii k\frac{\mathbf{R}^2}{2\Delta z}}.
\end{equation}
The propagation of a beam $b_{\mathrm{j}}(\mathbf{R})$ in a free space on a distance $\Delta z_\mathrm{j}$ is then expressed as
\begin{equation}
    b_{{\mathrm{j}+1}}(\mathbf{R})=\frac{-\ii k}{2\pi\Delta z_\mathrm{j}}\int\mathrm{d}^2\mathbf{R}' b_{\mathrm{j}}(\mathbf{R}')\,\ee^{\ii k\frac{\left(\mathbf{R}-\mathbf{R}'\right)^2}{2\Delta z_\mathrm{j}}},
    \label{eqn:planar}
\end{equation}
where $\Delta z_\mathrm{j}=z_{\mathrm{j}+1}-z_{\mathrm{j}}$.

\begin{figure}[h!]
        \centering
        \includegraphics[scale = 1]{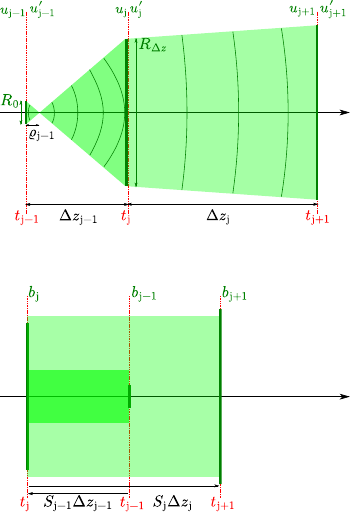}%
        \label{subfig:c}%
    \caption{
    A relation between the propagation of a beam function $b_\mathrm{j}(\mathbf{R})$ and $u_\mathrm{j}(\mathbf{R})=b_\mathrm{j}(\mathbf{R})\exp\{\ii k\mathbf{R}^2/(2\varrho_\mathrm{j})\}$.
    It turns out that $u_\mathrm{j}(\mathbf{R})$ being propagated on a distance $\Delta z_\mathrm{j}$ is completely equivalent to the beam $b_\mathrm{j}(M_{\rm j}\mathbf{R})$ being propagated on a distance $S_{\rm j}\Delta z_\mathrm{j}$,
    and then rescaled by the factor $M_\mathrm{j}=\sqrt{|\varrho_{\rm j}/(\Delta z_{\rm j}+\varrho_{\rm j})|}$ and multiplied by the phase factor corresponding to the parabolic wavefront.
    }
    \label{fig:Analytical}
\end{figure}

In this paper, we focus on the propagation of a beam through an optical system comprising various optical elements, each mathematically described by a transmission function $t_{\mathrm{j}}(\mathbf{R})$.
Let us assume that the wavefunction $u_{\mathrm{j}}(\mathbf{R})$ has a parabolic wavefront $\exp\{-\ii k\mathbf{R}^2/(2\varrho_{\mathrm{j}})\}$, with $\varrho_\mathrm{j}$ being the radius of curvature, in a plane of the $\mathrm{j}$-th optical element, as depicted in Fig.~\ref{fig:Analytical}.
Throughout this paper, we will use a convention that a diverging beam has a positive radius of curvature, whereas a converging beam has a negative radius of curvature.
The transmission through the element is then mathematically represented by a multiplication of $u_{\mathrm{j}}(\mathbf{R})$ with the respective transmission function $t_\mathrm{j}(\mathbf{R})$, so that the resulting wavefunction $u'_{\mathrm{j}}(\mathbf{R})$ can be written in a form
\begin{equation}
    u'_{\mathrm{j}}(\mathbf{R})=u_{\mathrm{j}}(\mathbf{R})t_{\mathrm{j}}(\mathbf{R})\,\equiv b_{\mathrm{j}}(\mathbf{R})\,\ee^{-\ii k\frac{\mathbf{R}^2}{2\varrho_\mathrm{j}}}.
\end{equation}
When propagated from the plane $z=z_\mathrm{j}$ to a plane $z=z_{\mathrm{j}+1}$, the resulting wavefunction $u_{{\mathrm{j}+1}}(\mathbf{R})$ will take the form
\begin{equation}
    u_{\mathrm{j}+1}(\mathbf{R})=\frac{-\ii k}{2\pi\Delta  z_{\mathrm{j}}}\int\mathrm{d}^2\mathbf{R}' b_{\mathrm{j}}(\mathbf{R}')\,\ee^{-\ii k\frac{\mathbf{R}'^2}{2\varrho_\mathrm{j}}}\ee^{\ii k\frac{\left(\mathbf{R}-\mathbf{R}'\right)^2}{2\Delta z_\mathrm{j}}},
\label{eqn:analyticalpropagation}
\end{equation}
when substituting $M_{\rm j}=\sqrt{|\varrho_{\rm j}/(\Delta z_{\rm j}+\varrho_{\rm j})|}$, $S_{\rm j}=\mathrm{sign}[\varrho_{\rm j}/(\Delta z_{\rm j}+\varrho_{\rm j})]$, $\mathbf{R}''=S_{\rm j}\mathbf{R}'/M_{\rm j}$, and rearranged,
\begin{equation}
\begin{aligned}
    u_{\mathrm{j}+1}(\mathbf{R})=S_{\rm j}M^2_{\rm j}\ee^{\ii k\frac{\mathbf{R}^2}{2(\varrho_\mathrm{j}+\Delta z_\mathrm{j})}}\frac{-\ii k}{2\pi S_{\rm j} \Delta z_\mathrm{j}}\int\mathrm{d}^2\mathbf{R}' b_{\mathrm{j}}\left(S_{\rm j}M_{\rm j}\mathbf{R}''\right)\times\\\times\exp{\left[\ii \frac{k}{2S_{\rm j}\Delta z_\mathrm{j}}\left(M_{\rm j}\mathbf{R}-\mathbf{R}'\right)^2\right]}.
\end{aligned}
\label{eqn:spherical}
\end{equation}
One can notice that wave $u_{{\mathrm{j}+1}}(\mathbf{R})$, given by Eq.~(\ref{eqn:spherical}), is of the same form as Eq.~(\ref{eqn:planar}) in the sense that it corresponds to a convolution of the beam $b_{\mathrm{j}}(\mathbf{R})$ and the Fresnel propagator.
The obvious difference is that the transverse coordinates of both the beam $b_{\rm j}(\mathbf{R})$ the resulting wavefunction $u_{\mathrm{j}+1}(\mathbf{R})$ are rescaled by a factors $S_{\rm j}M_\mathrm{j}$ and $M_\mathrm{j}$ respectively.
Another obvious difference is that the propagation distance $\Delta z_{\rm j}$ changes its sign when $\varrho_{\rm j}/(\Delta z_{\rm j}+\varrho_{\rm j})<0$, that is, when there is a beam crossover within the propagation interval. 
We can therefore claim that propagation of a wavefunction $u'_{\mathrm{j}}(\mathbf{R})$ from the plane of the $\mathrm{j}$-th optical element on a distance $\Delta z_\mathrm{j}$ is equivalent to propagation of $b_\mathrm{j}(S_{\rm j}M_\mathrm{j}\,\mathbf{R})$ in a free space on the distance $S_{\rm j}\Delta z_\mathrm{j}$, but the resulting wavefunction is multiplied by a factor $S_{\rm j}M^2_{\rm j}\exp\{{\ii k \mathbf{R}^2/[2(\varrho_\mathrm{j}+\Delta z_\mathrm{j})}]\}$ and its transverse coordinates are supposed to be rescaled by a factor $M_{\rm j}$. Putting it all together into Eq.~(\ref{eqn:spherical}) yields
\begin{equation}
    u_{\mathrm{j}+1}(\mathbf{R})=S_{\rm j}M^2_{\rm j}\,\ee^{\ii k\frac{\mathbf{R}^2}{2(\varrho_\mathrm{j}+\Delta z_\mathrm{j})}}\, b_{\mathrm{j}+1} \left(M_{\rm j}\mathbf{R}\right),
    \label{eqn:ekviv}
\end{equation}
where $b_{\mathrm{j}+1}(M_\mathrm{j}\mathbf{R})$ is $b_\mathrm{j}(S_{\rm j}M_{\rm j}\mathbf{R})$ being propagated on a distance $S_{\rm j}\Delta z_{\rm j}$ and then rescaled by factor $M_\mathrm{j}$.

Eq.~(\ref{eqn:ekviv}) shows that the propagated wavefunction $u_{\mathrm{j}+1}(\mathbf{R})$ has the same form as the original wave $u'_\mathrm{j}(\mathbf{R})$ in the sense that it has a form of product of some beam function and a parabolic wavefront.
Therefore, if any other optical element, given by a transmission function $t_{\mathrm{j}+1}(\mathbf{R})$, is present in a plane $z=z_{\mathrm{j}+1}$ and we wish to propagate it further, we can calculate the wavefunction $u_{\mathrm{j}+2}(\mathbf{R})$ simply by substituting $\varrho_{\mathrm{j}+1}=\varrho_\mathrm{j}+\Delta z_\mathrm{j}$, $M_\mathrm{j+1}=\sqrt{|\varrho_{\rm j+1}/(\Delta z_{\rm j+1} + \varrho_{\rm j+1})|}$ and $S_\mathrm{j+1}=\mathrm{sign}(\varrho_{\rm j+1}/(\Delta z_{\rm j+1} + \varrho_{\rm j+1}))$ into Eq.~(\ref{eqn:ekviv}). 
The desired function $b_{\mathrm{j}+2}\left(M_\mathrm{j+1}\mathbf{R}\right)$ is then calculated by propagating the function $S_{\rm j}M^2_\mathrm{j}\, b_{\mathrm{j}+1} \left(M_\mathrm{j}M_\mathrm{j+1}\mathbf{R}\right)\,t_{\mathrm{j+1}}(M_\mathrm{j+1}\mathbf{R})$ on a distance $S_{\rm j+1}\Delta z_{\rm j+1}$ using Eq.~(\ref{eqn:planar}) and then rescaling the resulting wavefunction by a factor $M_\mathrm{j+1}$.

The above derivation shows that wavefunction $u_\mathrm{j}(\mathbf{R})$ can be written in the form of Eq.~(\ref{eqn:ekviv}) anywhere in the optical system.
We can therefore apply this equivalence to any part of the optical setup or to the system as a whole.

Our proposed SASM is based on rescaling the propagation distances $\Delta z_\mathrm{j}$ and the radii of curvature $\varrho_\mathrm{j}$ in a limit $M_{\mathrm{j}}\rightarrow \infty$, i.e. in a vicinity of a point where the beam is focused. 
Then, if one compares wavefunctions $u_{\mathrm{j}+1}(\mathbf{R})$ (given by Eq.~(\ref{eqn:spherical})) and $\bar{u}_{\mathrm{j}+1}(\mathbf{R})$, which is also given by Eq.~(\ref{eqn:spherical}) with substitutions $\Delta \bar{z}_{\mathrm{j}}=\delta_\mathrm{j}\Delta z_{\mathrm{j}}$ and $\bar{\varrho}_{\mathrm{j}}=\delta_\mathrm{j} \varrho_{\mathrm{j}}$, then one obtains the following relation in a limit $M_\mathrm{j}\rightarrow \infty$
\begin{equation}
    \bar{u}_{\mathrm{j}+1}(\mathbf{R})=\frac{1}{\delta_\mathrm{j}}\ee^{\ii k\frac{\mathbf{R}^2}{2\delta_\mathrm{j}\Delta z_\mathrm{j}}\left(1-\frac{1}{\delta_\mathrm{j}}\right)}u_{\mathrm{j}+1}\left(\frac{\mathbf{R}}{\delta_\mathrm{j}}\right).
    \label{eqn:pom1}
\end{equation}
If $\delta_\mathrm{j}\gg1$, Eq.~(\ref{eqn:pom1}) can be simplified.
\begin{equation}
    \bar{u}_{\mathrm{j}+1}(\mathbf{R})=\frac{1}{\delta_\mathrm{j}}\ee^{\ii k\frac{\mathbf{R}^2}{2\delta_\mathrm{j}\Delta z_\mathrm{j}}}u_{\mathrm{j}+1}\left(\frac{\mathbf{R}}{\delta_\mathrm{j}}\right).
    \label{eqn:zdenek}
\end{equation}


The value of the scaling parameter $\delta_\mathrm{j}$  can be chosen freely, as long as we meet a condition Eq.~\eqref{Eq:Deltalimit} that limits its maximal value (see a detailed discussion in Sec.~\ref{Sec:Comparison}).
As discussed in the introduction, the larger $\delta_\mathrm{j}$, the faster the calculation will be.

\section{Numerical examples}
We demonstrate the use of the SASM and NLASM on two examples of numerical simulations for a simple one-lens system in Fig.~\ref{fig:FIG1}, and for a more complex setup containing two lenses and a phase plate in Fig.~\ref{fig:FIG2}. The simulations are tailored for electron wave optics, where the wave nature of high-energy electrons is modelled accurately, including propagation, diffraction, and aberration effects. We first describe the numerical propagation using the ASM, from which both SASM and NLASM are derived, and then explain how simulations using these two new methods are performed.

\subsection{Angular spectrum method}
At the beginning of each simulation, we define a complex wave function $u(\mathbf{R}, z_0)$ at an initial plane $z_0$. This wave is discretised on a two-dimensional grid and represented by a complex-valued matrix $\mathcal{M}_{\text{real space}} \in \mathbb{C}^{N \times N}$, where each matrix element corresponds to a pixel containing amplitude and phase information. To model the free-space propagation, we transform the wave function into the reciprocal space by a fast Fourier transform (FFT):

\begin{align}
\mathcal{M}_{\text{reciprocal space}} = \mathcal{F}\lbrace\mathcal{M}_{\text{real space}}\rbrace.
\label{Eq:M1}
\end{align}
This transformation yields the angular spectrum representation of the wave, where each pixel corresponds to a plane-wave component with specific transverse wavevector components. The wave is then propagated along the optical axis by multiplying each component by a phase factor

\begin{align}
\mathcal{M}'_{\text{reciprocal space}} = \mathcal{M}_{\text{reciprocal space}}\,  \exp(\ii K_z\Delta z).
\label{Eq:M2}
\end{align}
After this step, the propagated wave in real space is obtained via an inverse FFT

\begin{align}
\mathcal{M}'_{\text{real space}} = \mathcal{F}^{-1}\lbrace\mathcal{M}'_{\text{reciprocal space}}\rbrace
\label{Eq:M3}
\end{align}
We note that Eqs.~\eqref{Eq:M1}--\eqref{Eq:M3} are numerical equivalents of the analytical solution in Eq.~\eqref{eqn:FSPM}, expressed in three steps.

Besides the propagation in free space, we also take into account the action of electron-optics elements. We first introduce lenses, which we model as thin phase objects acting in real space and imposing a position-dependent phase shift that is incorporated by multiplying the beam profile by the transmission function $t_\mathrm{j}(\mathbf{R})=\exp(\ii \Delta\varphi)$. 
%
%
For a lens including defocus and third-order spherical aberration 
(whose strength is expressed by coefficients $C_{10}$, and $C_{30}$, respectively), and for a general 
working distance (cf.\ Eq.~\eqref{Eq:lenssimple}), the phase shift is

\begin{align}
\Delta \varphi &=\underbrace{-k \left(  \sqrt{R^2+f_{\text{col}}^2}-f_{\text{col}}\right)}_{\Delta\varphi_\mathrm{col}}  \underbrace{-k\left(\sqrt{R^2+\text{wd}^2}-\text{wd}\right)}_{\Delta\varphi_\mathrm{foc}} \nonumber\\
&\underbrace{-k \left(\frac{\, C_{10}}{\text{wd}} \left( \sqrt{R^2+\text{wd}^2}-\text{wd} \right)+ \frac{C_{30} R^4}{4 f^4} \right)}_{\Delta\varphi_\mathrm{def, sph}}
\label{Eq:Unscaled_long}
\end{align}
where we split the contributions expressing the collimating action ($\Delta\varphi_\mathrm{col}$), focusing ($\Delta\varphi_\mathrm{foc}$) and the defocus and spherical aberrations ($\Delta\varphi_\mathrm{def, sph}$), $f$ is the focal length, $f_{\text{col}}$ is the distance of the beam crossover before the lens (for a collimated beam, $f_{\text{col}}=\infty$).
The working distance $\mathrm{wd}$ (i.e.\ the distance from the lens to the focal plane)  is related to the focal length by
\begin{equation}
\mathrm{wd} = \frac{1}{1/f - 1/f_{\text{col}}}.
\label{Eq:wd}
\end{equation}
We note that Eq.~\eqref{Eq:Unscaled_long} can be easily modified by introducing other types of aberrations present in electron lenses.

\subsection{Scaling angular spectrum method}

The ``pure'' ASM introduced above serves as our base
computational framework. We now extend it by introducing the scaling ASM (SASM), which operates in the scaled coordinate system characterised by the scaling factor $\delta$.

Within the scaled formalism, the propagated wave field can be directly evaluated 
only in the focal plane. This restriction follows from Eqs.~\eqref{eqn:pom1} 
and \eqref{eqn:zdenek}, which rely on the assumption $M \rightarrow \infty$, and which is satisfied exclusively at the focal plane. Direct evaluation of planes displaced from focus would violate this assumption.

To clarify the implementation of SASM, it is convenient to conceptually decompose 
the phase shift imposed by the lens in Eq.~\eqref{Eq:Unscaled_long} as if it were produced by two infinitesimally separated thin phase elements. The first element comprises the terms $\Delta\varphi_\mathrm{col}$ (the beam collimation) and $\Delta\varphi_\mathrm{def,sph}$, which remain unchanged under scaling. On the other hand, the ``focusing'' term in Eq.~\eqref{Eq:Unscaled_long}, $\Delta\varphi_\mathrm{foc}$, is responsible for focusing the beam to the working distance $\mathrm{wd}$ 
and is directly connected to the condition $M \rightarrow 0$. 
Within SASM, only this term is modified by enlarging the working distance 
by the scaling factor $\delta$. The scaled phase shift therefore becomes
\begin{align}
\Delta \varphi_\mathrm{scaled} &=\Delta\varphi_\mathrm{col}  \underbrace{-k\left(\sqrt{R^2+\bar{\mathrm{wd}}^2}-\bar{\mathrm{wd}}\right)}_{\Delta\varphi_\mathrm{foc, scaled}} +\Delta\varphi_\mathrm{def, sph},
\label{Eq:Scaled_long}
\end{align}
%
where
\begin{equation}
\bar{\mathrm{wd}} = \delta \, \mathrm{wd}.
\end{equation}
Eq.~\eqref{Eq:Scaled_long} differs from 
Eq.~\eqref{Eq:Unscaled_long} only in the second term, which now focuses 
the beam to the extended working distance $\bar{\mathrm{wd}}$. 
This modification relaxes the sampling requirements 
(see Sec.~\ref{Sec:Comparison}) while preserving all other lens effects, 
including defocus and higher-order aberrations.

Since the scaled formalism allows direct evaluation only in the focal plane, 
wave fields at nearby planes are obtained by introducing controlled defocus. 
Physically, defocus shifts the effective observation plane while maintaining 
the validity of the scaling assumptions. The defocus is implemented by adjusting the coefficient $C_{10}$, which parametrises axial displacement from the focal plane. To obtain the wave field at a plane 
located a distance $\zeta$ from the focal plane ($z=\mathrm{wd}$), 
the defocus coefficient must be chosen as
\begin{equation}
C_{10} 
= 
\frac{\zeta \, \mathrm{wd}}{\mathrm{wd} + \zeta}
\approx 
\zeta,
\end{equation}
where the approximation holds for $\zeta \ll \mathrm{wd}$. In this way, controlled defocus provides a consistent method for probing planes near the focus without violating the assumptions of the propagation model.

As soon as we get the propagated wave from the scaled system $\bar{u}_{C_{10}}$ (with a defocus), we scale it along the $x$ and $y$ axes by $\delta$ and also multiply its amplitude by $\delta$ to match the corresponding wave $u$ in the original system [see Eq.~\eqref{eqn:pom1}]

\begin{align}
u(\mathbf{R}, z=z_{\text{focal plane}}+C_{10}) 	\approx \delta \, \bar{u}_{C_{10}}\left(\delta \mathbf{R}\right) \; \ee^{\Delta \varphi _{foc}}.
\label{Eq:ScaledWaveReconstruction}
\end{align}
Where $\ee^{\Delta \varphi _{foc}}$ is a phase shift representing a divergence or convergence of the wave $u$, which is missing in the collimated wave $ \bar{u}$.

\begin{figure*}
    \centering
    \includegraphics[width=0.9\textwidth]{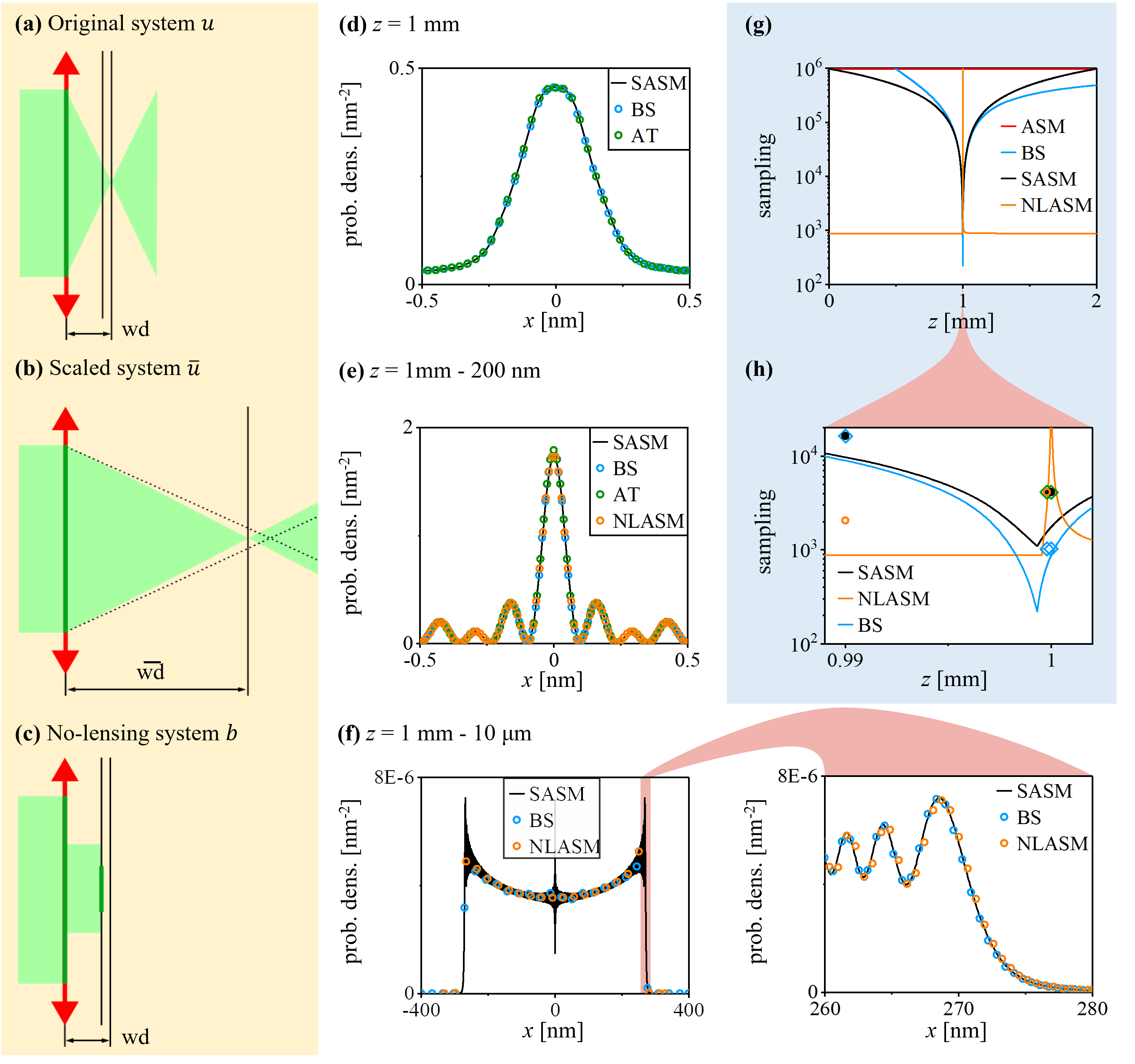}
    \caption{Numerical simulations of a wavefront propagated through an optical system consisting of one lens. \textbf{(a--c)} Schemes of original, scaled, and no-lensing optical systems. The model parameters are listed in Table~\ref {tab:SingleLens}.  wd is the working distance in the original system (1 mm), which is here equal to the focal distance of the lens, thanks to the collimated incoming beam. $\bar{\text{wd}}$ is the scaled working distance. \textbf{(d--f)} The beam intensity profile in planes 0 nm, $-200$~nm, and $-10\,\upmu$m from the working plane, respectively. The beam intensity is expressed as a probability density. The intensity profiles are calculated using various methods: the scaling angular spectrum method (SASM), the no-lensing angular spectrum method (NLASM), and the Bluestein (BS) method. For comparison, we also show profiles calculated by the abTEM code (AT)~\cite{Susi2020} by direct calculation of the angular spectra in the plane of interest, without propagation from the focusing lens. We zoom a detailed view on the edge of the beam in (f), which is an important part of the beam profile carrying information about the numerical aperture and defining the focused spot size. \textbf{(g, h)} The minimal necessary number of pixels in one dimension of the initial plane ($z=0$) in dependence on the $z$ coordinate of the plane of interest. ASM stands for the ``pure'' angular spectrum method. Discretely placed symbols in \textbf{(h)} show the sampling used to calculate the beam profiles shown in \textbf{(d--f)}. The green marks stand for AT.} 
    \label{fig:FIG1}
\end{figure*} 

We first demonstrate the SASM on a setup consisting of one lens with a spherical aberration sketched in Fig.~\ref{fig:FIG1}(a). The wave propagates from left to right. Initially, we have a fully defined, collimated top-hat beam. The beam is then focused by the lens into the working plane. The purpose of this simulation is to find a beam profile in the working plane and its vicinity. To facilitate this calculation for the computer according to the SASM, we can scale the system as shown in Fig.~\ref{fig:FIG1}(b).

Fig.~\ref{fig:FIG1}(d--f) shows a comparison of results acquired for the original (unscaled) system calculated using Bluestein and abTEM (BS, AT) approaches and the scaled system (SASM). We perform the calculations both in the focal plane $z=1$~mm (d) and at two defocused distances (e,f). The beam in the focal plane forms a small spot determined by the diffraction limit and spherical aberration. On the other hand, the defocused beams in (e,f) have more complicated profiles with side peaks. The detail in (f) shows an edge of the beam. In all three cases (d--f), we see that the profiles extracted from the scaled system are almost identical to those calculated without scaling, with approximately 2~\% difference as a result of remaining computational artefacts, which can be further reduced by increased sampling. The difference is calculated from the plotted probability density ($P$) of each method in Fig.~\ref{fig:FIG1}(d--f) with BS as its baseline:  

\begin{equation}
    \text{difference} = \frac{\int |P(x,y=0) - P_{\text{BS}}(x,y=0)| \; \mathrm{d}x}{\int  P_{\text{BS}}(x,y=0) \; \mathrm{d}x}.
\end{equation}

\subsection{No-lensing angular spectrum method}

\begin{figure*}
    \centering
    \includegraphics[width=0.9\textwidth]{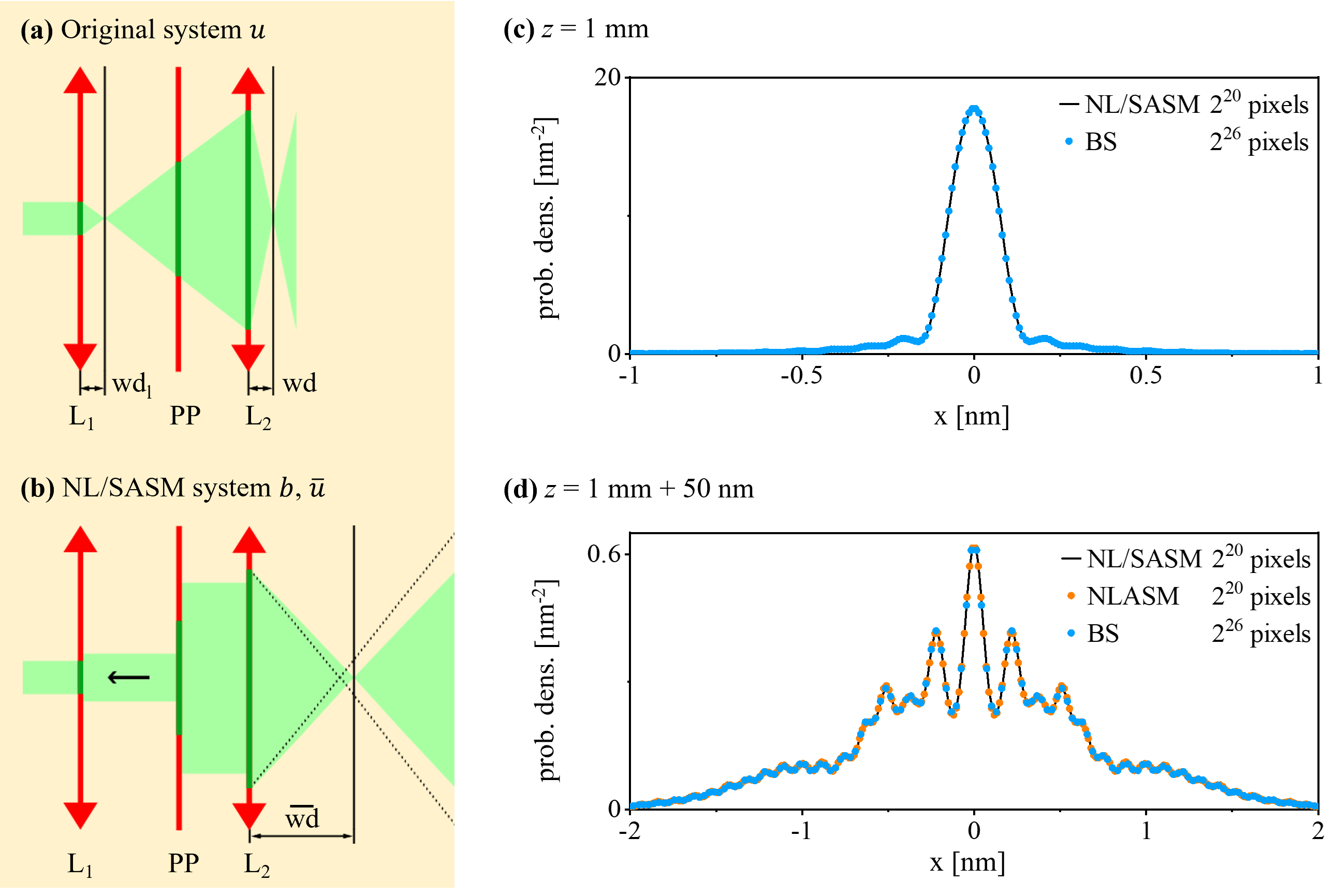}
    \caption{Numerical example of wave propagation through a complex optical system with multiple elements. \textbf{(a, b)} Schemes of original and no-lensing/scaled optical system. The model parameters are listed in Table~\ref {tab:PP}. wd$_l$ is a local working distance, wd is the final working distance, L$_1$ and L$_2$ are the first and second lenses, and PP is a phase plate. $\bar{\text{wd}}$ is the scaled final working distance. The beam undergoes a crossover in the section between L$_1$ and PP. The crossover in the scaled system is expressed as a reversed propagation direction in the L$_1$-PP section. \textbf{(c)} The beam intensity profile in the working plane, calculated by a combination of NLASM and SASM (black), and by BS (blue). \textbf{(d)} The beam intensity profile 50 nm behind the working plane, calculated by a combination of NLASM and SASM (black), only by NLASM (orange), and by BS (blue). The beam intensity is expressed as a probability density. The simulations carried out by NL/SASM, and NLSASM used 2$^{10}\times 2^{10} =$ 2$^{20}$ pixels. The simulations carried out by BS used $2^{13}\times2^{13}=2^{26}$ pixels.
    } 
    \label{fig:FIG2}
\end{figure*} 

While the SASM has been proven efficient for propagating waves to the focal plane or its vicinity, we have developed another method, the no-lensing angular spectrum method (NLASM), suitable for propagation towards planes far from the focal plane. The NLASM is based on completely removing the lens's focusing effect in a simulated setup and operates with collimated beams. We illustrate our approach with a specific example in Fig.~\ref{fig:FIG2}. The system consists of two thin lenses which are supposed to be scaled. Lens~1 (L$_1$) is assumed to be free of aberrations, whereas Lens~2 (L$_2$) exhibits spherical aberration characterized by a coefficient $C_{30} = 1\,\mathrm{mm}$. A thin phase plate (PP) is placed between the two lenses to compensate for the spherical aberration of L$_2$. The purpose of this simulation is to find a beam profile in the working plane of L$_2$, and its vicinity. To reach this goal, let us split this task into three steps: propagate the wave i) from  L$_1$ to PP, ii) from  PP to L$_2$, iii) from L$_2$ to the working plane.

The first step is to propagate the wave from L$_1$ to the phase plate (PP). 
We start from a fully defined, collimated top-hat wave illuminating L$_1$, 
and aim to determine the wave field incident on the PP. 
In this stage, we completely remove the focusing action of L$_1$, 
i.e., we omit the first and second terms of the polynomial in 
Eq.~\eqref{Eq:Scaled_long}. 
In the present case, since L$_1$ is aberration-free and no defocus is applied, 
the lens does not modify the beam in the no-lensing system. 
The wave is therefore propagated as a collimated beam over a distance 
$\Delta z$ from L$_1$ to the PP. 
This propagation distance is taken to be the same as in the original system.


To reproduce at the PP the same wavefront profile as in the original setup, we rescale the radius of the collimated beam to a value $\bar{R}$ at the beginning of the L$_1$–PP section:

\begin{align}
\bar{R} =\frac{R_0}{M}=R_0\sqrt{\left|1+\frac{\Delta z}{\varrho}\right|}=\sqrt{\frac{R_{\Delta z}}{R_0}}\,R_0 =\sqrt{R_0 R_{\Delta z}},
\label{Eq:ScaledColimatedBeam}
\end{align}
where $R_0$ is the beam radius at the beginning, and $R_{\Delta z}$ is the beam radius at the end of the propagation distance. The third equality can be deduced from Fig.~\ref{fig:Analytical} by applying a triangle similarity. Both $R_0$ and $R_{\Delta z}$ are always positive, therefore, $\bar{R} \in  \mathbb{R}^+$.

In Fig.~\ref{fig:FIG2}(a), we see a crossover in the L$_1$–PP section. To imitate the crossover with the collimated beam in the no-lensing system, we take the propagation distance $\Delta z$ in Eq.~\eqref{Eq:M2} as negative. This was already explained in the Analytical theory section. Generally, in the no-lensing system, all sections that imitate a crossover have a negative propagation distance, and all sections without a crossover have a positive propagation distance: 

\begin{align}
    \Delta z \in \Bigg\{
    \begin{array}{ll}
        \mathbb{R}^-, & \text{if crossover}\\
         \mathbb{R}^+, & \text{if no crossover}.
    \end{array}
\end{align}

After propagating from the starting to the ending point (i.e., from L$_1$ to PP), we scale the propagated beam again, by multiplying its transversal coordinates by a factor $\pm R_{\Delta z}/\bar{R}$, to match the radius of the original beam at PP. In the case of a crossover, we use ``$-$'' sign to simulate flipping the wave as it passes the crossover. In the case of no crossover, we use ``$+$'' sign. 

As soon as the wave is modified by the PP (which modulates the phase of the incoming wave), we again continue with the collimated beam. We have to scale the beam radius using Eq.~\eqref{Eq:ScaledColimatedBeam}. After propagation from PP to L$_2$, we scale the propagated beam again to match the original beam radius at L$_2$. The final step is to propagate the wave from L$_2$ to the working distance, using the same approach as for the propagation from L$_1$.

The results are shown in Fig.~\ref{fig:FIG2}(c, d). In (c), we see an electron beam profile in the focal plane, determined by the diffraction limit and also the fifth-order spherical aberration, which is a typical consequence of positioning the corrector (PP) out of the objective (L$_{2}$) plane~\cite{Beck2012}. In (d), we see a more complicated beam profile which is formed behind the focal plane, revealing an effect of the fifth-order spherical aberration. So far, it has been computationally difficult to reach such data without BS and simulate full beam profiles depending on the positioning and parameters of individual components in an electron microscope. NLASM and SASM make the simulation even easier, as they reduce the number of pixels by 64 times compared to BS in this example. The NLASM was also used for the comparison with other methods in Fig.~\ref{fig:FIG1}(e, f). Again, we see a good match to the results calculated by alternative methods, with a difference of around 2~\%.

\section{Comparison of the propagation methods}
\label{Sec:Comparison}
The individual methods introduced in this work require different sampling. Let us compare the methods in terms of the minimum number of pixels along a single dimension.

In the BS method, each pixel in the initial plane acts as a point in a diffraction grid. The grid creates a diffraction pattern in the final plane where we propagate the wave. It creates a periodic net of repeating beam profiles. Here, the key condition for the proper functionality of the method is that the individual profiles do not overlap. The denser the sampling in the initial plane, the bigger the distance between repeating beam profiles. For instance, in the case of a STEM, the beam is formed into a cone with a tip in the focal plane. The BS method creates a net of such cones that can overlap as we move farther from the focal plane. To avoid the overlap, the sampling needs to be increased. The distance between the centers of neighbouring periodic profiles $\Omega$ is 
\begin{align}
\Omega = \mathrm{tg}(\theta) z = \frac{\lambda z}{d \sqrt{1-\left(\frac{\lambda}{d}\right)^2 }} \approx  \frac{\lambda z}{d},
\end{align}
where we assumed $d\gg\lambda$. $z$ is the distance from the lens to the plane of interest. The minimal number of pixels along one dimension in the BS method can then be expressed as
\begin{align}
N_\mathrm{BS} > \frac{2 R_0 D}{\lambda z},
\label{Eq:N_BS}
\end{align}
where $R_0$ is the radius of the beam in the initial plane, and $D$ is the beam width in the plane of interest. For a perfectly focused beam without any aberrations, we can easily calculate the beam width in any plane $z$ and calculate the required $N_\mathrm{BS}$ as a function of the distance along the optical axis as 
\begin{align}
N_\mathrm{BS} > \frac{4 R^2}{\lambda}\frac{ |z - f|}{z f}.
\end{align}
However, we note that it is also necessary to consider other factors that determine the beam width $D$, such as diffraction and lens aberrations, which make the beam wider, especially in the vicinity of its focal plane.

In the ASM method, we have to fulfil the Nyquist criterion, which applies as keeping the phase difference between neighboring pixels below $\pi$:

\begin{align}
    N_\mathrm{ASM} > \frac{2k}{\pi} \left( \frac{R^2}{f} + \frac{C_{30}R^4}{f^4} \right).
    \label{Eq:NASM}
\end{align}

In the case of NLASM, the first term in brackets above, which corresponds to the lensing effect, drops out, and we obtain
\begin{align}
    N_\mathrm{NLASM} > \text{max} \Bigg\{
    \begin{matrix}
        \frac{2k}{\pi} \frac{C_{30}R^4}{f^4}\\
         \frac{2k}{\pi} \frac{C_{30}R^4}{f^4} \frac{D}{D_0},
    \end{matrix}
\end{align}
where we took into account the possibility of an additional padding since the spherical aberration can broaden the beam, and we do not want to hit the boundaries. The beam broadening is expressed by the ratio $D/D_0$, where $D_0$ is the width of a beam without the spherical aberration.

In the case of SASM, we can reduce the first term in the brackets of Eq.~\eqref{Eq:NASM} by the scaling factor $\delta$. On the other hand, the scaling will also bring the boundaries closer. But boundaries have to extend the beam width. This determines the maximum value of $\delta$ factor:
\begin{align}
    \delta_\mathrm{max} < \frac{2R}{D},
    \label{Eq:Deltalimit}
\end{align}
and thus
\begin{align}
    N_\mathrm{SASM} > 
    \frac{2k}{\pi} \left( \frac{R^2}{f \delta_\mathrm{max}} + \frac{C_{30}R^4}{f^4} \right) =
    \frac{2k}{\pi} \left( \frac{D R}{2f} + \frac{C_{30}R^4}{f^4} \right).
\end{align}

Figs.~\ref{fig:FIG1}(g, h) show a comparison of the minimal required pixels for sampling along one dimension for a wave propagation from an objective lens aberrated by the spherical aberration as a function of the propagation distance. For various distances, different methods are efficient. The BS method is the most efficient for propagating to the focal plane, but struggles with planes far from it. The SASM follows the BS method and can serve as a comparable alternative. The NLASM cannot provide results in the focal plane, where the sampling requirements jump to infinity, but it is very efficient in other regions. This makes the NLASM a complementary method to the BS and SASM. A combination of these three methods can simulate the electron wavefront propagation in any region of a conventional electron microscope.

\section{Summary and conclusion}
The proposed scaling angular spectrum method (SASM) and no-lensing angular spectrum method (NLASM) are novel numerical approaches for propagating complex waves through an optical setup, derived from the ASM. The SASM substitutes an original optical setup with a scaled setup, where the lenses have longer focal distances. Therefore, the propagated wave features much smaller phase gradients (the convergent and/or divergent angles are smaller), and the scaled system requires much sparser spatial sampling. We demonstrate that computational requirements for the computer's random-access memory and also the computational time are reduced by the factor $\delta^2$. In electron microscopy, the usual value of the scaling factor square is $\delta^2 = 10^6$. We also emphasise that SASM and NLASM are formulated without relying on the paraxial approximation. The NLASM reduces the sampling requirements by completely suppressing the lensing effect of simulated lenses and substitutes the original system with one with collimated beams.

The described methods for the wave propagation (BS, SASM, NLASM) can be well combined with particle tracing (PT) simulations. PT can work very well with thick optical elements of various shapes ~\cite{LENCOVA2008315}. Therefore, PT can be used for the initial analysis of an inspected electron-optical system to calculate parameters of individual components, such as lens focal length and aberration coefficients, which allows representing the optical system by thin elements and enables the use of the presented methods for efficient wave propagation.


The BS, SASM, and NLASM provide significant acceleration for simulations of complex wave propagation, which can be particularly crucial when designing electron optics and implementing novel elements, for instance, amplitude- and phase-modulating plates~\cite{shiloh2018, Shiloh_2019, Nekula2025, Mihaila2022, ChiritaMihaila2025, Ferrari2025}. It can also be crucial for detailed simulations and for understanding signals measured when using both standard and novel methods, such as 4D-STEM~\cite{Kayla2024, Nellist1998}.

Although simulations considering light optics have much less stringent sampling requirements, the same approach can be applied to efficiently compute advanced optical beam profiles~\cite{Allen1992_OAM, Forbes2021}. As a propagating electromagnetic wave is a vector field, we can use the described SASM and NLASM only in the paraxial regime, where the vector field can be approximated by a scalar field.

\section*{Acknowledgements}
We acknowledge the support of the Czech Science Foundation GACR under the Junior Star grant 23-05119M and the Ministry of Education, Youth and Sports of the Czech Republic under the project LL2506.

\appendix

\clearpage
\section{Parameters used in numerical calculations} 

\begin{table*}[h!]
\caption{
Single lens system in Fig.~\ref{fig:FIG1}
}
\begin{tabular}{lllll}
\hline\hline
& \textbf{BS}           & \textbf{SASM}  & \textbf{NLASM} & \textbf{AT}   \\ \hline\hline
Wavelength & $3.7014$ pm & $3.7014$ pm & $3.7014$ pm & $3.7014$ pm \\
Initial intensity distribution & tophat & tophat & tophat & tophat \\
Initial beam diameter & $60$ $\upmu$m & $60$ $\upmu$m & $60$ $\upmu$m & $60$ $\upmu$m \\
Focal length nominal $f$ & 1 mm & 1 mm & 1 mm & 1 mm \\
Spherical aberration $C_{30}$ & $1$ mm & $1$ mm & $1$ mm & $1$ mm \\ 
Convergence semiangle nominal & 30 mrad & 30 mrad & 30 mrad & 30 mrad\\
\hline
$z = 1$ \\
\hline
Scaling $\delta$ factor & - & $1000$ & - & - \\
Number of pixels in initial plane & 1024$\times$1024 & 4096$\times$4096 & - & 4096$\times$4096 \\
Difference to BS & - & 0.7 \% & - & 0.9 \% \\
\hline
$z = 1$ - 200 nm \\
\hline
Scaling $\delta$ factor & - & $$5000$$ & - & - \\
Number of pixels in initial plane & 1024$\times$1024 & 4096$\times$4096 & 4096$\times$4096 & 4096$\times$4096 \\
Difference to BS & - & 1.3 \% & 1.4 \% & 2.1 \% \\
\hline
$z = 1$ - 10 $\mu$m \\
\hline
Scaling $\delta$ factor & - & $$80$$ & - & - \\
Number of pixels in initial plane & 16384$\times$16384 & 16384$\times$16384 & 2048$\times$2048 & - \\
Difference to BS & - & 2.2 \% & 2.2 \% & - \\
\hline\hline       
\end{tabular}
\label{tab:SingleLens}
\end{table*}

\begin{table*}[h!]
\caption{
Two-lens system in Fig.~\ref{fig:FIG2}
}
\begin{tabular}{llll}
\hline\hline
& \textbf{BS}  & \textbf{NL/SASM} & \textbf{NLASM}               \\ \hline\hline
Space width / beam width &  2 & 2 & 2 \\
Wavelength & $3.7014$ pm & $3.7014$ pm & $3.7014$ pm \\
Initial intensity distribution & tophat & tophat & tophat\\
Number of pixels in space & $8192 \times 8192$ & $1024 \times 1024$ & $1024 \times 1024$ \\
Initial beam radius & $2$ $\upmu$m & $2$ $\upmu$m & $2$ $\upmu$m \\
L$_1$ to PP beam radius & - & $4.472$ $\upmu$m & $4.472$ $\upmu$m   \\
PP to L$_2$ beam radius & - & $14.142$  $\upmu$m & $14.142$  $\upmu$m  \\
L$_1$ focal length nominal $f$ & 10 mm & 10 mm  & 10 mm \\
L$_2$ focal length nominal $f$ & 0.99 mm & 0.99 mm & 0.99 mm \\
L$_2$ C$_{30}$ & 1 mm & 1 mm & 1 mm\\
Working distance wd & 1 mm & 1 mm & 1 mm \\
Working distance scaled $\bar{\text{wd}}$ & -  & 10 m & - \\
Convergence semiangle nominal & 20 mrad & 20 mrad & 20 mrad\\ 
L$_1$ to L$_2$ distance & 110 mm & 110 mm & 110 mm \\
PP to L$_2$ distance & 50 mm & 50 mm & 50 mm\\
PP C$_{30}$ & $-1$~mm & $-1$~mm & $-1$~mm \\
\hline  
$z = 1$ mm \\
\hline  
Scaling $\delta$ factor & -  & $10000$ & - \\ 
Difference to BS & - & 0.6 \%  & - \\
\hline  
$z = 1$ mm + 50 nm \\
\hline  
Scaling $\delta$ factor & - & $5000$  & - \\ 
Difference to BS & - & 0.7 \% & 1.4\%  \\

\hline\hline       
\end{tabular}
\label{tab:PP}
\end{table*}

\clearpage

\bibliographystyle{elsarticle-harv} 
\bibliography{main}






\end{document}